\documentclass[aps, pre,reprint,superscriptaddress]{revtex4-1}
\usepackage{graphicx}

\newcommand{\eq}[1]{(\ref{#1})}
\newcommand{\pd}{\partial}

\begin{document}
\title{Kinetic simulations of ladder climbing by electron plasma waves}
\author{Kentaro Hara}
\email[Electronic email: ]{khara@tamu.edu}
\affiliation{Department of Aerospace Engineering, Texas A\&M University, College Station, Texas, 77843, USA}
\affiliation{Princeton Plasma Physics Laboratory, Princeton, New Jersey, 08543, USA}
\author{Ido Barth}
\affiliation{Department of Astrophysical Sciences, Princeton University, Princeton, New Jersey, 08544, USA}
\affiliation{Racah Institute of Physics, Hebrew University of Jerusalem, Jerusalem 91904, Israel}
\author{Erez Kaminski}
\affiliation{Birmingham-Southern College,  Birmingham, Alabama, 35254, USA}
\altaffiliation{Currently at Wolfram Research Inc., 100 Trade Center Drive, Champaign, Illinois, 61820, USA.}
\author{I. Y. Dodin}
\affiliation{Princeton Plasma Physics Laboratory, Princeton, New Jersey, 08543, USA}
\affiliation{Department of Astrophysical Sciences, Princeton University, Princeton, New Jersey, 08544, USA}
\author{N. J. Fisch}
\affiliation{Princeton Plasma Physics Laboratory, Princeton, New Jersey, 08543, USA}
\affiliation{Department of Astrophysical Sciences, Princeton University, Princeton, New Jersey, 08544, USA}

\begin{abstract}
The energy of plasma waves can be moved up and down the spectrum using chirped modulations of plasma  parameters, which can be driven by external fields. Depending on whether the wave spectrum is discrete (bounded plasma) or continuous (boundless plasma), this phenomenon is called ladder climbing (LC) or autoresonant acceleration of plasmons. It was first proposed by Barth \textit{et al.} [PRL \textbf{115}, 075001 (2015)] based on a linear fluid model. In this paper, LC of electron plasma waves is investigated using fully nonlinear Vlasov-Poisson simulations of collisionless bounded plasma. It is shown that, in agreement with the basic theory, plasmons survive substantial transformations of the spectrum and are destroyed only when their wave numbers become large enough to trigger Landau damping. Since nonlinear effects decrease the damping rate, LC is even more efficient when practiced on structures like quasiperiodic Bernstein-Greene-Kruskal (BGK) waves rather than on Langmuir waves \textit{per~se}.
\end{abstract}

\maketitle


\section{Introduction}

Ladder climbing (LC) is understood as an approach to a robust excitation of quantum systems by the means of chirped quasiperiodic modulation of system parameters. Such modulation, or drive, induces successive Landau-Zener (LZ) transitions~\cite{landau32, zener32} between neighboring energy levels when the corresponding transition frequency is in resonance with the drive. As the modulation is chirped, transitions are induced in different pairs of levels at different times. Then it becomes possible to robustly propel quanta across a wide range of the energy spectrum, provided that the chirp rate is slow enough and the drive is sufficiently strong.

By now, LC has been demonstrated in various quantum systems ranging from atoms and molecules~\cite{chelkowski95, maas98, marcus04, marcus06}, to anharmonic oscillators~\cite{barth11, barth13, barth14}, Josephson resonator~\cite{shalibo12}, and bouncing neutrons~\cite{manfredi17}. In the limit of continuous spectrum, the drive couples many levels simultaneously and the quantum LC become the well known classical autoresonance (AR)~\cite{meerson90, fajans99, fajans01, ben-david06, barak09, friedland06, khain07, barth08}. Most recently, it was also proposed that the effect is extendable to classical systems~\cite{barth15}. Specifically, it was shown in Ref.~\citenum{barth15} that Langmuir waves in bounded plasma may undergo LC much like a quantum system, if the background plasma density is subjected to a low-frequency chirped modulation (e.g., a chirped acoustic wave). However, the theory in Ref.~\citenum{barth15} relies on a linear fluid model, so it neglects kinetic effects, such as Landau damping, and nonlinear effects, such as particle trapping. Whether LC by electron plasma waves can survive these effects and can be practiced on realistic waves remains to be shown \textit{ab~initio}.

The purpose of this paper is to present first \textit{ab~initio} collisionless simulations confirming that LC of electron plasma waves is a robust effect that can survive kinetic and nonlinear effects. The simulations are done using a one-dimensional Vlasov-Poisson code. We find that, at sufficiently low mode numbers numbers $m$, LC proceeds much like anticipated from the simplified fluid theory~\cite{barth15}. At larger $m$, Landau damping and nonlinear effects eventually disrupt the process. That said, we also find that nonlinear effects facilitate LC in the sense that they reduce Landau damping and thus help plasmons reach $m$ larger than those expected from the linear theory. In other words, LC is even more efficient when practiced on quasiperiodic Bernstein-Greene-Kruskal (BGK) modes~\cite{bgk57, dodin14book} rather than on linear waves \textit{per~se}. 

The LC phenomenon practiced upon plasma waves is certainly of academic interest, because the Langmuir wave is probably the most fundamental and widely occurring mode in plasma physics. However, manipulating its properties through ladder climbing could be of interest in practical applications as well. Certain applications exploit the small group velocity of the Langmuir wave, such as plasma holography~\cite{dodin02}, plasma photonic crystals~\cite{lehmann16}, and other cooperative plasma phenomena~\cite{rousseaux16}. The plasma wave is also useful in mediating the compression of laser energy in plasmas, thereby to reach ultra-high intensities~\cite{malkin99}. In that regard, the ability of the plasma wave to linger in plasma owing to its small group velocity makes it a useful seed for this interaction~\cite{qu17}. In each of these cases, while the plasma wave is lingering, but before performing a task, such as, retrieving information or mediating laser compression, it can be imagined that it might be usefully manipulated to better perform that task. The LC described here would be one tool to perform those manipulations or optimizations.

The paper is organized as follows. In Sec.~\ref{sec:basic}, we briefly overview the fluid theory reported in Ref.~\citenum{barth15}. In Sec.~\ref{sec:kinetic}, we introduce our numerical model. In Sec.~\ref{sec:results}, we produce our main results. In Sec.~\ref{sec:conc}, we present our main conclusions.

\section{Fluid theory of ladder climbing}
\label{sec:basic}
Consider a one-dimensional collisionless nonmagnetized plasma with immobile ions that form a static homogeneous background. As known commonly from fluid theory~\cite{stixbook}, such plasma supports electrostatic electron waves, called Langmuir waves, whose frequency $\omega$ for a given wave number $k$ is given by $\omega = \omega_{pe} [1 + 3 (k \lambda_D)^2]^{1/2}$. Here, $\omega_{pe} = (4 \pi n_0  e^2/m_e)^{1/2}$ is the electron plasma frequency, $\lambda_D =  v_{{\rm th},e} / \omega_{pe}$  is the Debye length, $n_0$ is the unperturbed electron density, $e$ is the elementary charge, $m_e$ is the electron mass, $v_{{\rm th},e} =(T_e /m_e)^{1/2}$ is the electron thermal velocity, and $T_e$ is the electron temperature. 

Assuming hard-wall boundary conditions, the allowed wave numbers are $k_m = m k_1$, where $m$ is the mode number, $k_1 = \pi/L$ is the wave number of the fundamental mode, and $L$ is the plasma length. The discrete dispersion relation of a standing Langmuir wave can be written as  $\omega_m \approx \omega_{pe} (1 + \tilde{\beta} m^2)^{1/2}$, where $\tilde{\beta} = 3 \pi^2 \lambda_D^2/L^2$. Note that $\tilde{\beta}$ can be understood as a measure of the spectrum anharmonicity, i.e., of how strongly the frequency difference of neighboring modes $\omega_{m,m+1} = \omega_{m+1} - \omega_{m}$ depends on $m$. For $\tilde{\beta}  m^2\ll 1$, one has
\begin{equation}\label{eq:omm}
\frac{\omega_{m,m+1}}{\omega_{pe}} = \sqrt{1+\tilde{\beta} (m+1)^2} - \sqrt{1+\tilde{\beta} m^2} \approx  \tilde{\beta} \left(m+\frac{1}{2}\right).
\end{equation} 
As any collection of discrete modes, such system is mathematically equivalent to a quantum particle governed by a Hamiltonian with the same spectrum~\cite{dodin14}. Thus, linear Langmuir waves in bounded fluid plasma can be described by LC theory borrowed from quantum mechanics, in which the system  is propelled  from an initial mode {(e.g., the lowest-order mode, or ``ground state'')} up to a desired final mode~\cite{barth11, barth15}. LC can be realized by applying an external drive or  a density modulation~\cite{barth15}, with a chirped frequency  $\omega_d = \omega_{0} + \alpha t$, where $\omega_0$ is the starting frequency, $\alpha$ is a constant chirping rate, and $t$ is time.

Following the quantum LC theory, we identify two dimensionless parameters of interest: the driving parameter $P_1 = A/(4 \sqrt{\tilde{\alpha}})$ and the anharmonicity parameter $P_2 = \tilde{\beta} /\sqrt{\tilde{\alpha}}$, where $A$ is the modulation amplitude (namely, the relative perturbation of the background electron density) and $\tilde{\alpha} = \alpha/\omega_{pe}^2$ is the dimensionless chirping rate. The probability of the plasmon transfer between neighboring modes is given by $1 - \exp(-\pi P_1^2 /2)$~\cite{landau32,zener32}. In order to have efficient LC, $P_1$ must be large enough. For example, $P_1 > 1.5$ results in energy transfer above 97\% to the next mode. In addition, from Eq.~(\ref{eq:omm}), one has $\omega_{m,m+1} - \omega_{m-1,m} \approx \omega_{pe}\tilde{\beta}$. This means that the time interval between successive resonances (``transition time'') is $\Delta t_{\rm trans}= \omega_{pe}\tilde{\beta}/\alpha = \omega_{pe}^{-1}\tilde{\beta}/\tilde{\alpha}$. Using the ``natural" dimensionless time $\tau = \sqrt{\alpha} t = \sqrt{\tilde{\alpha}} \omega_{pe} t$, the transition time is given by   $\Delta \tau_{\rm trans} = \tilde{\beta} / \sqrt{\tilde{\alpha}} = P_2$. For LC, $P_2 \gg 1 + P_1$ must be satisfied so that the LZ transitions are well separated and only two levels are coupled at a given time. In the other limit, where $P_2 \ll 1$, many levels are simultaneously coupled and the system exhibits AR acceleration, which is the continuum limit of LC. Also note that $\tilde{\alpha} \ll 1$ (adiabaticity condition) is needed for this theory to hold. Otherwise, the mode coupling induced by the drive is nonresonant, so the transfer of quanta becomes phase-dependent {(nonadiabatic)}.

This theory of LC and AR by Langmuir waves was proposed in Ref.~\citenum{barth15}, and it was also confirmed there numerically using linear fluid simulations. Although the linear Landau damping was recognized as a kinetic limit on the accessibility of levels with high $m$, the kinetic stability of lower levels and the phase space evolution during the damping were not studied.  In order to explore how LC is modified when kinetic and nonlinear effects are involved, more rigorous simulations are needed. We report such simulations below. The transition to the AR is not considered because of numerical limitations.

\section{Kinetic model }
\label{sec:kinetic}
Electrons are described by their phase-space distribution $f$, which is a function of the position $x$, velocity $v$, and time~$t$. We adopt the reflecting-wall conditions in $x$ space; i.e., $f(x, v, t) = f(x, -v, t)$ at the plasma boundaries $x = 0$ and $x = L$. The dynamics of $f$ is governed by the Vlasov equation
\begin{equation}
\label{eq:vlasov}
\frac{\partial f}{\partial t} + v \frac{\partial f}{\partial x} - \frac{e E}{m_e}  \frac{\partial f}{\partial v} = 0,
\end{equation}
where the electric field is given by $E = E_{\rm s} + E_{\rm drive} + E_{\rm prep}$, where $E_{\rm s}$ is the self-induced field, $E_{\rm drive}$ is the  field that drives LC, and the ``preparation" field $E_{\rm prep}$ is  used to set up the initial Langmuir wave. The self-induced field is $E_{\rm s} = - \pd_x\phi$, where the potential $\phi$ is governed by the Poisson equation
\begin{equation}
\label{eq:possion}
\frac{\partial^2 \phi}{\partial x^2} = - 4\pi e (n_i - n_e).
\end{equation}
Here, $n_i$ is the ion density, which is constant (in both $x$ and $t$), and $n_e(x, t) = \int^\infty_{-\infty} f(x, v, t)\,dv$ is the electron density, respectively. We assume that the plasma is overall neutral [$\int_0^L n_e(x, t)\,dx / L = n_i$] and the surface charges at the walls are zero, so the boundary conditions for the electric field are $E(x = 0) = E(x = L) = 0$.

In this paper, we investigate the LC dynamics {that begins the ``ground level''; namely, the initial wave is prepared using $E_{\rm prep}$ that is resonant with the lowest mode ($m=1$). We adopt} $E_{\rm prep}(x,t') = E_{{\rm p}0} \hat{A}_{p}(t') \cos(\omega_1 t') \sin(k_1 x)$, where $E_{{\rm p}0} $ is the amplitude of the  preparation driver, $t' = t - t_0$, and $t_0$ is the starting time of the simulation. Following Refs.~\citenum{berger13} and \citenum{banks11}, we choose a ramp-up and ramp-down envelope as follows: $2 \hat{A}_{p}(t') = \tanh [8 (t'/t_r-0.5)]-\tanh[8 ((t'-t_c)/t_r-0.5)]$.  The time scale of the ramp-up and ramp-down stages, $t_r$, is chosen large enough to prevent beating of the plasma wave with the preparation  field and thus retain a smooth distribution; specifically, we choose $t_r = 40 \omega_{pe}^{-1}$. The time $t_c$ during which the   amplitude is kept constant is chosen to be $t_c = 200\omega_{pe}^{-1}$. It is noted that the initial wave action in the first mode depends on the preparation field amplitude, $E_{{\rm p}0}$, and duration, $t_c$. 

After the initial mode is excited, we turn off $E_{\rm prep}$ and apply a different, chirped external field
\begin{equation}
E_{\rm drive}(x,t) = E_{{\rm d}0} \sin(k_1 x) \cos \left(\omega_{\text{1,2}} t + \alpha \frac{t^2}{2}\right).
\end{equation}
The frequency of this field, $\omega_d = \omega_{\text{1,2}} + \alpha t$, is initially in resonance with the frequency of the transition between the first and second modes, $\omega_{\text{1,2}}=\omega_2 - \omega_1$. Note that $\omega_{1,2} \ll \omega_{pe}$ for $\tilde{\beta}m \ll 1$ [see Eq.~\eq{eq:omm}] and  $t=0$ is chosen to be the time when $\omega_d = \omega_{\text{1,2}}$. At later times, $\omega_d$ becomes resonant with the transition frequencies $\omega_{m,m+1}$ corresponding to higher $m$, so plasmons can be gradually propelled from the lowest mode to higher modes, thus realizing LC.

In order to have efficient LC, the values of $\tilde{\alpha}$ and $\tilde{\beta}$ are chosen based on the following conditions. First, the system length must satisfy $L / \lambda_D  = \pi / (k_1  \lambda_D) =  (3/\tilde{\beta})^{1/2} \gg 3 \pi $ in order to ensure that kinetic effects are weak ($k_m \lambda_D \lesssim 1/3$)~\cite{berger13} at least for the first few resonant modes ($m \sim 1$).  Thus, $\tilde{\beta} \ll 1/(3 \pi^2).$ Second, $P_2 =\tilde{\beta}/\sqrt{\tilde{\alpha}}\gg 1 + P_1$ {is adopted to ensure the LC regime (see Sec.~\ref{sec:basic}).} For the  simulations reported here, we chose $\tilde{\alpha}=4.5628 \times 10^{-8}$ and $\tilde{\beta} = 0.002$. These parameters correspond to $L / \lambda_D \approx 121$ and $\Delta t_{\rm trans} = 4.38 \times 10^4 \omega_{pe}^{-1}$, i.e., $\tau_{\rm trans} = P_2 = 9.36$.   In addition, we employ $E_{{\rm p}0} = 1 \times 10^{-4}$ and $E_{{\rm d}0} = 0.1$, which yields $P_1 \approx 2.4$ in our simulation, for which the transition probability predicted from fluid theory is almost 100\%.

The numerical method chosen to solve Eq.~\eq{eq:vlasov} is Strang's time splitting with a finite volume method using the monotonic upwind for scalar conservation laws (MUSCL) scheme~\cite{vanleer4}. A modified Arora-Roe limiter~\cite{arora97} is used in order to preserve positivity of the phase-space distribution $f$ and reduce the numerical dissipation as much as possible within the MUSCL framework. Since simulations were done for large time scales (about $10^5$ plasma periods),  Message Passing Interface (MPI) is used for parallel computing. Previously, this method was applied for simulating plasma discharges in Hall thrusters~\cite{harajap14}, trapped particle instability~\cite{hara15}, and plasma wall interactions~\cite{hanquist17}.  The computational time for one simulation is about 1-2 days using 64 processors. The resolution of the Vlasov simulation is set as follows: $\Delta x = L/N_x \approx \lambda_D / 5$ and $\Delta v = (v_{\rm max}-v_{\rm min})/N_v$, where $N_x = 512$, $v_{\rm max} = -v_{\rm min} =  8 v_{{\rm th},e}$, $N_v=1000$, and $v_{{\rm th},e}$ is the electron thermal velocity. The time step is $ \Delta t \approx 0.028\omega_{pe}^{-1}$, the total steps $N_t = 2.7 \times 10^7$, resulting in the total time about $7.7 \times 10^5 \omega_{pe}^{-1}$.

\begin{figure}
\includegraphics[width=0.48\textwidth]{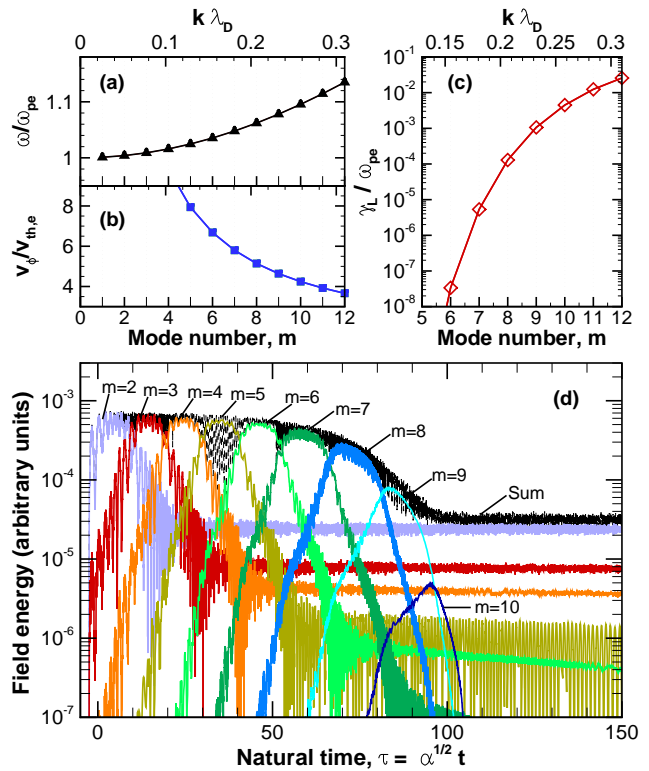}
\caption{\label{fig:1}
Simulated evolution of plasma waves during LC: (a) wave frequency, $\omega$; (b) phase velocity, $v_\phi$; (c) linear Landau damping rate, $\gamma_L$, from Eq.~\eq{eq:landau}; and (d) the evolution of the field-energy spectrum.  Also shown is the sum of the field energy for $m \ge 2$.	The contribution of mode with $m=1$ is excluded in order to eliminate the interference with $E_{\rm drive}$, which has the wave number equal to that of the first mode. }
\end{figure}

\section{Results}
\label{sec:results}

\begin{figure}[t]
\includegraphics[width=0.45\textwidth]{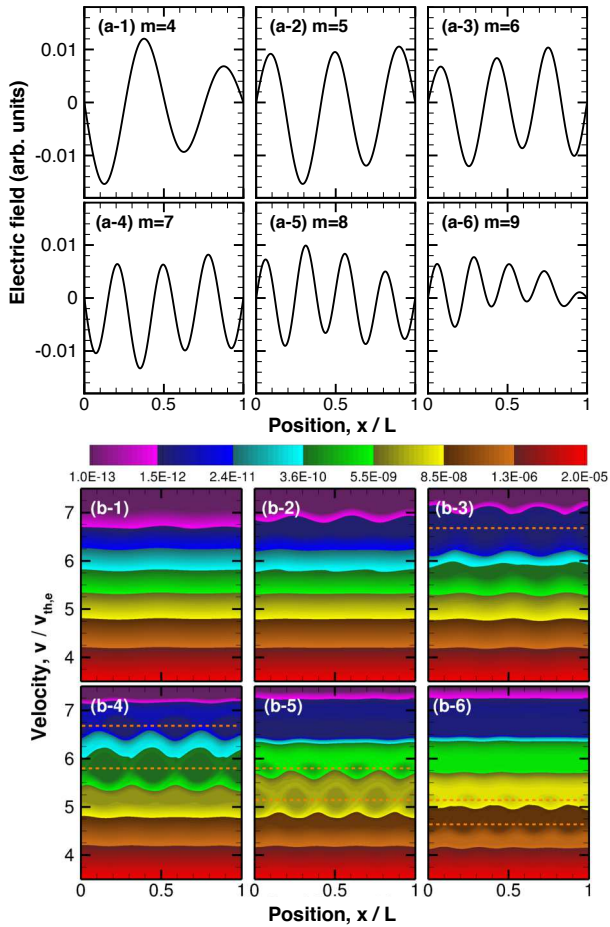}
\caption{\label{fig:2} 
Snapshots of a plasma wave undergoing LC: (a) electric field; and (b) electron distribution at $\tau =$ 21.1, 33.9, 43.1, 55.9, 70.6, and 85.2, when transitions occur to modes with $m=4, 5, 6, 7, 8$, and 9 (see Fig.~\ref{fig:1}). The same log-scale colormap is used in all six subfigures in (b). The orange dashed lines in (b) show the linear phase velocities of the relevant modes. }
\end{figure}

\textit{Field spectrum.} Figure~\ref{fig:1} presents an overview of the electron plasma wave evolution. Figures~\ref{fig:1}(a)-(c) show the wave frequency $\omega$, phase velocity $v_\phi = \omega/k$, and linear Landau damping rate $\gamma$, respectively, as functions of the mode number $m$ and the corresponding wave number $k$ (i.e., $k_m$). The real part of the frequency is calculated using the {fluid dispersion of Langmuir wave (Sec.~\ref{sec:basic})}, and the Landau damping is calculated using~\cite{stixbook}
{
\begin{equation}
\label{eq:landau}
\gamma_L \approx \sqrt{\frac{\pi}{8}}\, \frac{\omega_{pe}}{(k \lambda_D)^3}\, \exp\left[ - \frac{1}{2 (k\lambda_D)^2} - \frac{3}{2}\right].
\end{equation}
}
At $m \lesssim 7$, Landau damping is negligible at our parameters, so the wave total action $I$ is conserved~\cite{dodin14}. {Since the Langmuir wave temporal spectrum is localized in the vicinity of $\omega_{pe}$, one can adopt the standard linear relation between the action and the wave energy $\mathcal{E}$, namely, $I \approx \mathcal{E}/\omega_{pe}$~\cite{dodin09}. At small enough $k\lambda_D$, one also has $\mathcal{E} \approx 2W$, where $W = \int_0^L E^2/(8\pi)\,dx$ } is the total field energy~\cite{dodin09}. Then, $W$ is approximately conserved too. At larger $m$, this approximation fails, and, eventually, the action conservation is also broken, namely, due to Landau damping. This evolution is illustrated in Fig.~\ref{fig:1}(d). Specifically, we plot {$W_m = \int_0^L E_m^2/(8\pi) dx$}, where $E_m$ is the amplitude of the spatial mode with the corresponding $m$ calculated using Fourier decomposition, $E_m= ({2}/L) \int_0^L E \sin(k_m x) dx$. Also note that the transitions between individual modes from the numerical simulation occur at multiples of time periods which are predicted from fluid theory up to $m=5$, $\Delta \tau_{\rm trans} = 9.36$ ($\Delta t _{\rm trans}= 4.38 \times 10^4 \omega_{pe}^{-1}$ in the true dimensional time) for our simulation. At $m \gtrsim 5$, the transition time becomes larger than what the fluid theory predicts, because kinetic corrections to the wave dispersion relation becomes substantial. Below, we discuss some aspects of kinetic effects in more detail.

\textit{Particle distribution.} The characteristic temporal evolution of a plasma wave during LC is shown in Fig.~\ref{fig:2}. The snapshots illustrating oscillations at modes with $m = $ 4, 5, 6, 7, 8, and 9 correspond to $\tau =$ 21.1, 33.9, 43.1, 55.9, 70.6, and 85.2 in Fig.~\ref{fig:1}(d), respectively. As plasmons get transferred to higher and higher $m$, the phase velocity of the wave decreases and approaches the bulk in the distribution function [Fig.~\ref{fig:1}(b)]. Modes with $m > 4$ carry a noticeable amount of trapped electrons, but the real part of the frequency is largely unaffected by the trapped population. This is seen in Fig.~\ref{fig:2}(b) that shows the corresponding distribution functions and $v_\phi/v_{{\rm th},e}$ calculated from the linear theory.

\begin{figure}[tb]
\includegraphics[width=240pt]{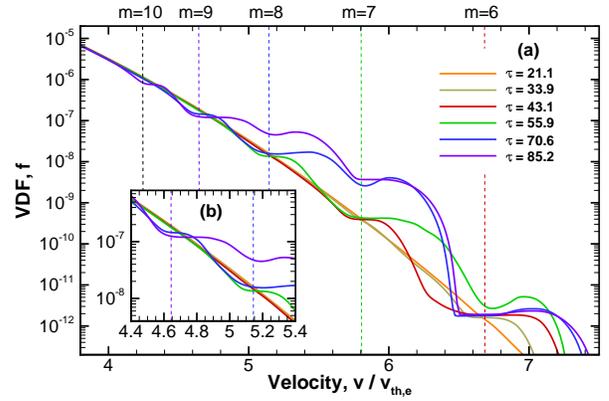}
\caption{\label{fig:1d}
Spatially-averaged electron velocity distribution functions (VDFs). The time steps correspond to the ones shown in Fig.~\ref{fig:2}. Plateaus of the VDFs are found around the phase velocities predicted by the analytic theory. The insert, Fig.~\ref{fig:1d}(b), is a zoomed-in view showing the evolution of the modes with $m=8$ and $m=9$.
}
\end{figure}

Figure~\ref{fig:1d} shows the spatially-averaged electron velocity distributions (VDFs). Flattened VDFs are formed around the phase velocity predicted by the analytic theory. However, flattening of the spatially-averaged VDF of the next mode can be also seen. For instance, the analytic theory predicts that, at $\tau = 43.1$, plasma oscillations are excited at $m = 6$, which corresponds to $v_\phi/v_{{\rm th}, e} = 6.68$. However, Fig.~\ref{fig:1d} shows VDF flattening also around $v_\phi/v_{{\rm th},e} = 5.80$ (red line), which corresponds to $m=7$. This can be explained by the fact that LC is not an abrupt but rather continuous process. It can indeed be seen from Fig.~\ref{fig:2}(b-3) that particles around $v_\phi/v_{{\rm th},e}=5.80$ are modulated but not fully trapped as the potential amplitude of $m=7$ is still increasing. As seen in Fig.~\ref{fig:1}(d), the wave energy of the next mode increases exponentially before the transition occurs. This results in adiabatic trapping of particles around the phase velocity of the following mode. 

\begin{figure}[t!]
\includegraphics[width=240pt]{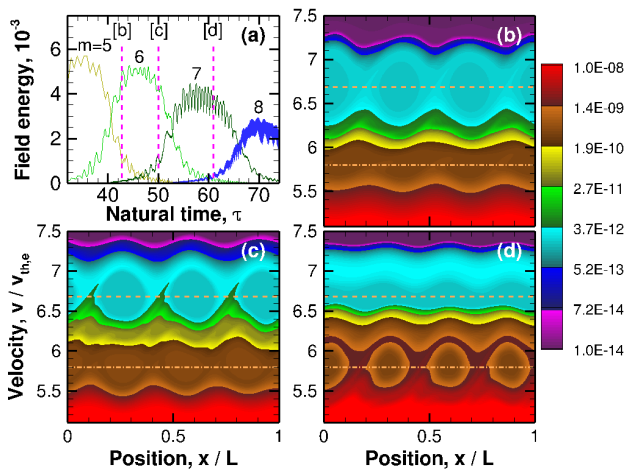}
\caption{\label{fig:3} 
An illustration of a transition between two modes, specifically $m=6$ to $m=7$: (a) wave energy, zoomed-in from Fig.~\ref{fig:1}. The zoomed-in electron distribution shown in (b), (c), and (d) correspond to the moments of time marked in (a) with pink dashed lines. The orange dashed lines in (b)-(d) show the phase velocities of the modes with $m=6$ and $m=7$ correspondingly.
}
\end{figure}

In detail, the transition from $m=6$ to $m=7$ can be seen in Fig.~\ref{fig:3}. Particle trapping occurs at $v_\phi / v_{{\rm th},e}= 6.68$, which corresponds to the $m=6$ mode. For a sinusoidal wave, the size of the trapped particle region is $\Delta v_{\rm tr} = 2\sqrt{e E_0/(k m_e)}$, where $E_0$ is the wave amplitude~\cite{berger13}.  {Due to the approximate energy conservation (see above), $E_0 \approx \text{const}$}.  Thus, $\Delta v_{\rm tr}$ decreases with $m$, and this effect is seen in simulations indeed [Fig.~\ref{fig:2}(b)]. The effect is strengthened by the fact that, at large $m$, Landau damping comes into play; then $E_0$ is not conserved but actually decreases too, as will be discussed below in detail.

We also performed simulations with other amplitudes of the seeded wave, which results in different $E_0$. Larger-amplitude plasma waves exhibit trends similar to those seen in Figs.~\ref{fig:1}-\ref{fig:3}. The main difference is that, at larger amplitudes, the size of the trapping islands increases, because $\Delta v_{{\rm tr}} \propto \sqrt{E_0}$. Eventually, $\Delta v_{{\rm tr}}$ exceeds the difference between the phase velocities of neighboring modes, which is given by (assuming $\tilde{\beta}m^2 \ll 1$)
\begin{equation}
\label{eq:vdiff}
v_{m+1}-v_m \approx \left(\frac{1}{m+1} -  \frac{1}{m}\right) \frac{\omega_{pe}}{k_1}  = - \frac{\omega_{pe}/k_1}{m(m+1)}.
\end{equation}
This causes nonlinear interactions between the modes. While a slight kinetic dissipation is observed, LC can still occur even when $\Delta v_{{\rm tr}}/2 < v_{m+1} - v_m$. The corresponding simulations are not presented in this paper.

\begin{figure}[t!]
\includegraphics[width=240pt]{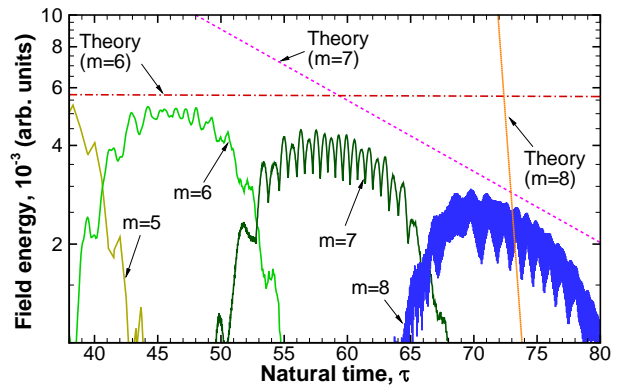}
\caption{\label{fig:4} 
Comparison between our numerical simulation and  the analytic theory for Landau damping. The dash-dotted, dashed, dotted lines are theoretical prediction of Landau damping for $m=6$, $m=7$, and $m=8$, respectively.
}
\end{figure}

\textit{Effect of Landau damping.} Figure~\ref{fig:4} compares predictions of Eq.~(\ref{eq:landau}) for the rate of linear Landau damping with numerical simulations. The Landau damping rate is too small to matter for modes with $m \le 6$. For $m = 7$, one can expect a 40\% energy loss to Landau damping during the transition time $\Delta \tau_{\rm trans}$. For $m = 8$, the linear theory predicts that the wave energy decreases during transition by orders of magnitude. Such strong dissipation is not observed in reality due to nonlinear effects, because we operate in the regime of relatively large bounce frequency $\omega_B = \sqrt{ekE_0/m_e}$.  The corresponding bounce period $t_B = 2 \pi/ \omega_B$ is about $120\omega_{pe}^{-1}$, which is much smaller than the transition time. Moreover, $\gamma_L t_B \ll 1$ for all modes of interest ($\gamma_L t_B \approx 5 \times 10^{-6}$, $7.9 \times 10^{-4}$, $0.018$, and $0.14$, for $m=6$, 7, 8, and 9, respectively). This implies that the modes are in the strongly nonlinear regime and are not Langmuir waves \textit{per~se}; rather, they can be considered as quasiperiodic BGK-like modes. Since nonlinear effects suppress Landau damping, they facilitate LC in the sense that they help plasmons reach higher $m$. But of course, at very large $m$, linear damping is still stronger than the nonlinearity, so there is a limit on the maximum $m$ (in our case $m \approx 9$) beyond which LC is impossible.

\begin{figure}[tb!]
\includegraphics[width=240pt]{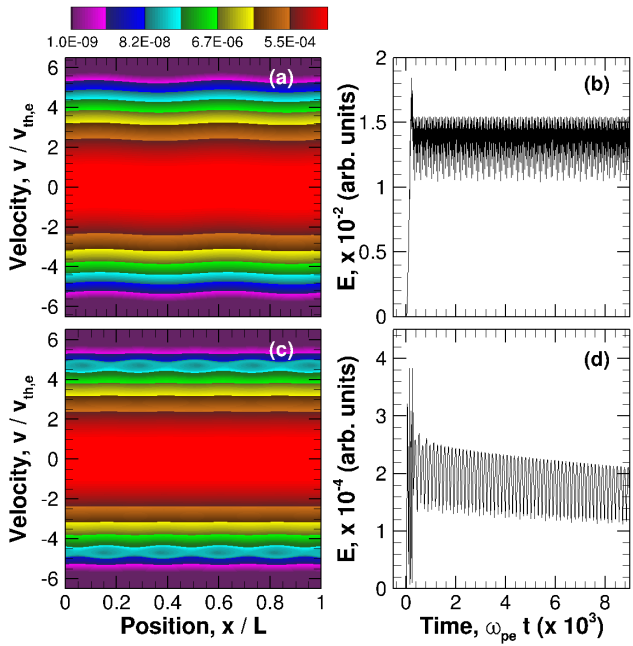}
\caption{\label{fig:5} 
Simulations of a single mode plasma wave with reflecting-wall boundary conditions. Figures (a, b) correspond to the mode with $m = 4$, and (c, d) correspond to the mode with $m = 9$. Particle trapping occurs around $v_\phi = \pm 4.64 v_{{\rm th},e}$.
}
\end{figure}

\textit{Kinetic dissipation of counter-propagating waves.} It is to be noted that, even in the absence of linear Landau damping, some nonlinear dissipation is always present in the system due to reflecting walls. This is due to the fact that a wave with a positive wave number is also accompanied by a wave with a negative wave number. In that case, there is no reference frame where the electric field would be stationary, so true BGK waves are impossible; i.e., no propagating structure is truly stationary. As pointed out earlier in Ref.~\citenum{schmit11}, there always remains some amount of interaction between nonlinear waves propagating in the opposite directions, resulting in dissipation. 

This effect is illustrated in Fig.~\ref{fig:5} that shows the evolution of two single modes, namely, with $m = 4$ and $m = 9$. The former has $v_\phi/v_{\rm th} \sim 10$, so it carries no trapped particles and is essentially linear; hence the amplitude of the field stays constant and the wave exhibits no damping. In contrast, the latter has $v_\phi/v_{\rm th} \sim 5$, so the trapped-particle content is noticeable. That makes the wave nonlinear, thus resulting in damping.

\section{Conclusions}
\label{sec:conc}

In summary, we report the first {\it ab initio} simulations of LC by electron plasma waves that was originally proposed in Ref.~\citenum{barth15} within a linear fluid theory. The simulations are done using a one-dimensional collisionless Vlasov-Poisson code. We find that, although the original theory was simplified, it does, in fact, capture the essential features of the phenomenon in realistic settings that involve both kinetic and nonlinear effects. Specifically, we find that, at sufficiently low mode numbers numbers $m$, LC is kinetically stable and is much like predicted in Ref.~\citenum{barth15}. At larger $m$, Landau damping and nonlinear effects eventually disrupt the process. That said, we also find that nonlinear effects facilitate LC in the sense that they somewhat suppress Landau damping due to particle-trapping and flattening of the distribution function and thus help plasmons reach $m$ larger than those expected from the linear theory. In other words, LC happens to be more efficient when practiced on BGK modes rather than on linear Langmuir waves \textit{per~se}. {Such modes are potentially producible in nonneutral-plasma experiments using Penning traps~\cite{danielson04} and are similar to driven phase space holes that can be excited autoresonantly using externally imposed standing waves~\cite{barth08}. (For boundless plasmas, a similar excitation technique using traveling waves was also reported in Refs.~\citenum{friedland06} and ~\citenum{khain07}.) It is to be noted that, although the LC dynamics of BGK-like modes is qualitatively discussed in this paper, a full kinetic theory of LZ-type transitions between such modes remains to be developed.}

\begin{acknowledgments}

The authors acknowledge L. Friedland for useful comments. The work was supported by the U.S. NNSA SSAA Program through DOE Research Grant No. DE-NA0002948, the U.S. DTRA Grant No. HDTRA1-11-1-0037, and the U.S. DOE through Contract No. DE-AC02-09CH11466. K.H. acknowledges the Japan Society for the Promotion of Sciences (JSPS) Postdoctoral Fellowship. 
\end{acknowledgments}


\begin{thebibliography}{10}
\bibitem{landau32}
L.~D. Landau, Phys. Z. Sowjetunion {\bf 2}, 46 (1932).

\bibitem{zener32}
C. Zener, Proc. R. Soc. A. {\bf 137}, 696 (1932).

\bibitem{chelkowski95}
S. Chelkowski and G.~N. Gibson, Phys. Rev. A {\bf 52}, R3417 (1995).

\bibitem{maas98}
D. Maas, D. Duncan, R. Vrijen, W. van~der Zande, and L. Noordam, Chem. Phys.
  Lett. {\bf 290}, 75 (1998).

\bibitem{marcus04}
G. Marcus, L. Friedland, and A. Zigler, Phys. Rev. A {\bf 69}, 013407 (2004).

\bibitem{marcus06}
G. Marcus, A. Zigler, and L. Friedland, Europhys. Lett. {\bf 74}, 43 (2006).

\bibitem{barth11}
I. Barth, L. Friedland, O. Gat, and A.~G. Shagalov, Phys. Rev. A {\bf 84},
  013837 (2011).

\bibitem{barth13}
I. Barth and L. Friedland, Phys. Rev. A {\bf 87}, 053420 (2013).

\bibitem{barth14}
I. Barth and L. Friedland, Phys. Rev. Lett. {\bf 113}, 040403 (2014).

\bibitem{shalibo12}
Y. Shalibo, Y. Rofe, I. Barth, L. Friedland, R. Bialczack, J.~M. Martinis, and
  N. Katz, Phys. Rev. Lett. {\bf 108}, 037701 (2012).

\bibitem{manfredi17}
G. Manfredi, O. Morandi, L. Friedland, T. Jenke, and H. Abele, Phys. Rev. D
  {\bf 95}, 025016 (2017).

\bibitem{meerson90}
B. Meerson and L. Friedland, Phys. Rev. A {\bf 41}, 5233 (1990).

\bibitem{fajans99}
J. Fajans, E. Gilson, and L. Friedland, Phys. Rev. Lett. {\bf 82}, 4444
  (1999).

\bibitem{fajans01}
J. Fajans and L. Friedland, Am. J. Phys. {\bf 69}, 1096 (2001).

\bibitem{ben-david06}
O. Ben-David, M. Assaf, J. Fineberg, and B. Meerson, Phys. Rev. Lett. {\bf 96},
  154503 (2006).

\bibitem{barak09}
A. Barak, Y. Lamhot, L. Friedland, and M. Segev, Phys. Rev. Lett. {\bf 103},
  123901 (2009).

\bibitem{friedland06}
L. Friedland, P. Khain, and A.~G. Shagalov, Phys. Rev. Lett. {\bf 96}, 225001
  (2006).

\bibitem{khain07}
P. Khain and L. Friedland, Phys. Plasmas {\bf 14}, 082110 (2007).

\bibitem{barth08}
I. Barth, L. Friedland, and A.~G. Shagalov, Phys. Plasmas {\bf 15}, 082110
  (2008).

\bibitem{barth15}
I. Barth, I.~Y. Dodin, and N.~J. Fisch, Phys. Rev. Lett. {\bf 115}, 075001
  (2015).

\bibitem{bgk57}
I.~B. Bernstein, J.~M. Greene, and M.~D. Kruskal, Phys. Rev. {\bf 108},
  546 (1957).

\bibitem{dodin14book}
I.~Y. Dodin, Fusion Sci. Tech. {\bf 65}, 54 (2014).

\bibitem{dodin02}
I.~Y. Dodin and N.~J. Fisch, Phys. Rev. Lett. {\bf 88}, 165001 (2002).

\bibitem{lehmann16}
G. Lehmann and K.~H. Spatschek, Phys. Rev. Lett. {\bf 116}, 225002 (2016).

\bibitem{rousseaux16}
C. Rousseaux, K. Glize, S.~D. Baton, L. Lancia, D. Benisti, and L. Gremillet,
  Phys. Rev. Lett. {\bf 117}, 015002 (2016).

\bibitem{malkin99}
V.~M. Malkin, G. Shvets, and N.~J. Fisch, Phys. Rev. Lett. {\bf 82}, 4448
  (1999).

\bibitem{qu17}
K.~Qu, I.~Barth, and N.~J. Fisch, arXiv1612.06450 (2017).

\bibitem{stixbook}
T.~H. Stix, {\it Waves in Plasmas\/} (AIP, New York, 1992).

\bibitem{dodin14}
I.~Y.~Dodin, Phys. Lett. A {\bf 378}, 1598 (2014).

\bibitem{berger13}
R.~L. Berger, S. Brunner, T. Chapman, L. Divol, C.~H. Still, and E.~J. Valeo,
  Phys. Plasmas. {\bf 20}, 032107 (2013).

\bibitem{banks11}
J.~W. Banks, R.~L. Berger, S. Brunner, B.~I. Cohen, and J.~A.~F. Hittinger,
  Phys. Plasmas {\bf 18}, 052102 (2011).

\bibitem{vanleer4}
B. {Van Leer}, J. Comp. Phys. {\bf 23}, 276 (1977).

\bibitem{arora97}
M. Arora and P.~L. Roe, J. Comp. Phys. {\bf 132}, 3 (1997).

\bibitem{harajap14}
K.~Hara, M.~J. Sekerak, I.~D. Boyd, and A.~D. Gallimore, J. Appl. Phys. {\bf
  115}, 203304 (2014).

\bibitem{hara15}
K.~Hara, T.~Chapman, J.~W. Banks, S. Brunner, I. Joseph, R.~L. Berger, and
  I.~D. Boyd, Phys. Plasmas {\bf 22}, 022104 (2015).

\bibitem{hanquist17}
K.~M. Hanquist, K. Hara, and I.~D. Boyd, J. Appl. Phys. {\bf 121}, 053302
  (2017).

\bibitem{dodin09}
I.~Y. Dodin, V.~I. Geyko, and N.~J. Fisch, Phys. Plasmas {\bf 16}, 112101
  (2009).

\bibitem{schmit11}
P.~F. Schmit, I.~Y. Dodin, and N.~J. Fisch, Phys. Plasmas {\bf 18}, 042103
  (2011).

\bibitem{danielson04}
J.~R. Danielson, F. Anderegg, and C.~F. Driscoll, Phys. Rev. Lett. {\bf 92},
  245003 (2004).
\end{thebibliography}
\end{document}